\documentclass[11pt,a4paper]{article}
\usepackage[a4paper,total={160mm,220mm}]{geometry}

\usepackage[dvipsnames]{xcolor}
\usepackage{jheppub}
\usepackage{graphicx}
\usepackage{epsfig}
\usepackage{dcolumn}
\usepackage{bm}
\usepackage{amsmath}
\usepackage{siunitx}
\usepackage{amssymb}
\usepackage{physics}
\usepackage{slashed}
\usepackage{dsfont}
\usepackage{subcaption}
\usepackage{hyperref}
\usepackage{cleveref}
\usepackage{soul}
\usepackage{verbatim}
\usepackage{tikz}
\usetikzlibrary{arrows.meta,positioning,calc}
\usepackage{mhchem}

\hypersetup{
	colorlinks=true,       
	linkcolor=purple,   
	citecolor=blue,        
	filecolor=magenta,      
	urlcolor=blue           
}

\graphicspath{{Figures/}}

\DeclareMathOperator{\sinc}{sinc}
\DeclareMathOperator{\sinhc}{sinhc}

\definecolor{mygreen}{RGB}{0,119,112}
\definecolor{myred}{RGB}{180,30,30}
\definecolor{myblue}{RGB}{0,68,130}
\definecolor{myyellow}{RGB}{200,160,0}

\arxivnumber{xxx}
\preprint{MPP-2026-69}

\title{Oscillating Imprints of Dark Matter in Mesons Decays}

\author[a,b]{Prisco Lo Chiatto}
\author[a]{Babette D{\"o}brich}
\author[b]{Gilad Perez}
\affiliation[a]{Max Planck Institute for Physics, Boltzmannstr. 8, 85748 Garching, Germany}
\affiliation[b]{Department of Particle Physics and Astrophysics, Weizmann Institute of Science, Rehovot, Israel
7610001}
\date{}

\emailAdd{prisco.lo.chiatto@mpp.mpg.de}
\emailAdd{babette@mpp.mpg.de}
\emailAdd{gilad.perez@weizmann.ac.il}

\makeatletter
\gdef\@fpheader{}
\makeatother

\abstract{ We study scenarios in which ultralight dark matter (ULDM) causes oscillations of the Cabibbo--Kobayashi--Maskawa (CKM) matrix elements, considering two frameworks.
The first, previously proposed in the literature, employs the Nelson--Barr mechanism to solve the strong CP problem and the CKM phase is identified with a pseudo-Nambu--Goldstone boson.
The second, inspired by Froggatt--Nielsen flavor models, relies on quadratic couplings of the ULDM to the Standard Model while naturally suppressing linear couplings.

\noindent On the experimental side, we outline a strategy to search for such oscillations at flavor factories using meson decays, focusing on the NA62 experiment as the most promising candidate for discovery thanks to its large kaon statistics.
We show that the sensitivity of lifetime-based observables is parametrically degraded when the total particle flux is not known exactly, leading to a substantial loss of sensitivity compared to naive estimates.
We therefore advocate alternative observables based on direct counting of events, which retain the expected $1/\sqrt{N}$ scaling and provide a robust probe of oscillating CKM elements.
Our results highlight flavor experiments as a novel probe of ULDM through time-dependent signatures.
}

\begin{document}
\maketitle
\section{Introduction}

The origin of dark matter (DM) is one of the most profound questions in physics. 
Arguably ultralight DM (ULDM) provides the simplest solution to this question via the misalignment mechanism~\cite{Preskill:1982cy,Abbott:1982af,Dine:1982ah}, a non-thermal production mechanism which does not require any interaction other than gravity and is effective for masses $\lesssim \SI{1}{eV}$.
Theoretically, ultralight scalars and in particular pseudo-scalars are well-motivated and ubiquitous in extensions of the Standard Model (SM).
These include, among others, the QCD axion~\cite{Preskill:1982cy,Abbott:1982af,Dine:1982ah} (see ref.~\cite{Yu:2023gdq,Dobrich:2025oso}, for recent reviews on the QCD axion as a DM candidate), the dilaton~\cite{Arvanitaki:2014faa} (see, however,~\cite{Hubisz:2024hyz}), relaxion models~\cite{Graham:2015ifn,Banerjee:2018xmn}, Higgs portals~\cite{Piazza:2010ye}, and alternatives to the QCD axion~\cite{Dine:2024bxv}. 

In most of the above models of ULDM the most prominent direct signatures can be divided into two main classes: pseudoscalars and scalars.
Pseudoscalars can be searched for via magnetometers, spinometers and a variety of detectors derived from the axion-EM modified Maxwell  theory~(see ref.~\cite{Adams:2022pbo} for a recent review).
Pure scalar models, on the other hand, lead to variation of fundamental constants; the behavior of a ULDM candidate $\phi$ is approximately that of a background field oscillating with an amplitude equal to $\sqrt{2\rho_{\rm DM}}/m_{\phi}$, where $\rho_{\rm DM}$ is the local DM density, at a frequency equal to its mass $m_{\phi}$.
The coupling of the ULDM field to SM operators then induces an effective, spacetime-dependent, variation of fundamental constants.
Scalar ULDM has been searched for indirectly in fifth-force experiments, as well as directly using clocks or tests of violation of equivalence principle (EP) (see ref.~\cite{Antypas:2022asj} for a recent review and refs.\ therein), with \ce{^{229}Th} spectroscopy yielding a new opportunity~\cite{Arakawa:2026mls}.

In ref.~\cite{Dine:2024bxv}, another type of ULDM phenomenology was proposed, associated with the Nelson--Barr solution of the strong CP problem.
It was noticed that if CP is embedded within an spontaneously broken approximate U(1) horizontal flavor symmetry, then the CKM phase can be identified to linear order with the resulting pseudo-Nambu--Goldstone boson (pNGB).
Various mechanisms are explored in ref.~\cite{Dine:2024bxv} to keep its mass in the ULDM range, providing a candidate DM particle with a new kind of signature.
Indeed, due to the ULDM oscillations, both the CKM phase and the CKM angles oscillates with time.
This allows for a search of ULDM at accelerator-based experiments, in contrast with the low energy probes mentioned above.\footnote{We also mention other studies of the constraints on the Nelson--Barr ULDM model~\cite{Dai:2025von,Guo:2026lzo}, as well as proposal for other accelerator-based searches of ULDM~\cite{Bigaran:2025uzn,Bauer:2026dgu,Fieg:2026zdr}.}
At higher loop level, the model also predicts coupling of the light scalar to the nucleons, which cause violations of the equivalence principle (EP), posing strong constraints on the models.
We further discuss this construction below and also present a new model which does not have such couplings at tree-level, leading to different phenomenology, by coupling to SM quadratically.

A crucial aspect of the linearly-coupled scenario is that viable parameter space requires the ULDM field to couple predominantly to the first two generations.
This follows because couplings to the third generation generically induce large effects proportional to the corresponding Yukawa couplings, leading to strong constraints from equivalence-principle tests and other precision probes.
As a result, the dominant observable effects are expected in transitions involving light quarks, and in particular in kaon and B meson physics.
This observation, combined with the much larger statistics available at kaon experiments, singles out the NA62 experiment as a particularly promising probe of this class of models.
In fact, a larger number of meson decays are recorded at NA62 than at any other current or planned flavor experiment, making it especially sensitive to time-dependent effects of the type considered here.
Focusing on the NA62 experiment forces us to confront the important question of how the large boost of kaons affects the sensitivity.
This point was raised in ref.~\cite{Dine:2024bxv} but not analysed in detail.
We will see that this invalidates the estimates in ref.~\cite{Dine:2024bxv} for NA62, but that a judicious choice of experimental observable can not only recover but even surpass the quoted sensitivity.

The structure of the remaining of this paper is as follows. 
In \cref{sec:model} we introduce the models leading to oscillation of the CKM elements. 
In \cref{sec:Lifetime} we discuss the effect of the CKM oscillation on the lifetime of mesons, as well as a generic strategy to measure time-oscillating observables. 
In \cref{sec:Obs} we discuss the optimal observable to be used.
In \cref{sec:NA62} we recount some facts of the NA62 experiment, and  we conclude in \cref{sec:Conclusions}. Three appendices complement the paper.

\section{Oscillating CKM from ULDM}
\label{sec:model}
In this part we introduce two models that lead to oscillating CKM elements. The first is based on the construction presented in ref.~\cite{Dine:2024bxv}. We show that already in the two generation limit, where the SM CKM phase is not physical, we do obtain oscillation of the CKM elements as well as CP violation.
The second construction shows that CKM oscillation can be naturally induced in models where the CKM elements and the induced coupling to other SM fundamental parameters occur at quadratic order, which removes the EP/fifth-force bounds from $\phi$ exchange.
Unfortunately, we will see that EP bounds are reintroduced if $\phi$ is dark matter~\cite{Hees:2018fpg}.
\subsection{Oscillating CKM from Nelson--Barr ULDM}
Consider the minimal model of ref.~\cite{Bento:1991ez}. It introduces an additional vector-like quark pair $q$ ($\bar q$) that carries the same (opposite) SM charge as the right-handed up-quark, in addition to a neutral complex scalar, $\Phi=(f+\rho)\exp(i\theta)/\sqrt2$\,. The Lagrangian contains the following couplings
\begin{equation}
    {\cal L} \supset \mu \bar q q + (g_{i} \Phi+\tilde g_{i}\Phi^*) \bar u_{i} q+ y^{u}_{ij} \tilde{H}Q_i \bar u_{j}+ y^{d}_{ij} H Q_i \bar d_{j} \, .
    \label{eq:lagrangian}
\end{equation}
It is further assumed that the theory is CP-conserving, and CP is only spontaneously broken by the expectation value of $\Phi$\,, such that $\mu,y^u$ and $y^d$ are real. 
In the presence of the Higgs vacuum expectation value, $v$\,, the up-quark mass is given by the following $4\times 4$ matrix:
\begin{equation}
    {\cal M}_u = \left ( \begin{matrix} \mu & B  \cr 0 & m_u \end{matrix} \right ),\ \ m_u = y^uv,~B_i=(g_{i} \Phi+\tilde g_{i}\Phi^*)\,.
    \label{eq: 4times4 quark matrix}
\end{equation}
Due to the absence of the bottom left entry, and the fact that $\Phi$ only appears in the off-diagonal entry of the above matrix, $\arg(\det({\cal M}_u))=0$ holds and no QCD phase is introduced (while the CKM phase is unconstrained).  This can be ensured by introducing an additional ${Z}_2$ symmetry under which $q,~\overline{q}$ and $\Phi$ are odd while the SM fields are even~\cite{Barr:1984qx, Nelson:1983zb}. Here we use instead an approximate U(1) flavor symmetry to enforce this structure while also protecting the mass of the phase of $\Phi$~\cite{Dine:2024bxv}, which is a pNGB denoted as $\phi$. This U(1) is non-anomalous, leading to an ameliorated quality problem.

To identify the quark masses and mixing angles, we focus on the structure of  ${\cal M}_u{\cal M}_u^\dagger$\,.
Assuming that the vector-like quark is heavy $\mu,|B|\gg m_u$\,, we can integrate it out and are left with an effective up-quark mass matrix $\tilde{m}_u$ satisfying
\begin{equation}
    (\tilde m_u \tilde m_u^\dagger )_{ij}=\left (  (m_u m_u^T )_{ij} - \frac{(m_u)_{ik} B^\dagger_k B_\ell (m_u^T)_{\ell j}} {\mu^2 + B_fB_f^\dagger} \right ) \,.
    \label{eq:EffMassMat}
\end{equation}
The CKM matrix is the product of the $SU(3)$ matrix required to diagonalize $\tilde m_u \tilde m_u^\dagger$ and the left $SO(3)$ rotation required to diagonalize $y^{d}$\,. Assuming $\langle\theta\rangle={\cal O}(1)$\,, $\mu\lesssim |B|$\,, and the vectors $g$ and $\tilde g$ are of comparable magnitude and not parallel in flavor space (see discussion in ref.~\cite{Davidi:2017gir}), the resulting CKM matrix has an $\mathcal{O}(1)$ CP violating phase.

 The couplings of $\phi$ are found by replacing the VEV $\theta$ with $\theta_{\rm CKM}/2+\phi/f$. We take $g\propto(1,0,0),\ \tilde{g}\propto (0,1,0)$\,, with $m^u$ diagonal and $m^d=V^{d\dagger}\mathrm{diag}(m_d,m_s,m_b)$\,, where $V^{d}\in SO(3)$ is the real-valued CKM matrix of the original Lagrangian. 

These parameters can arise naturally, for example if a global shift symmetry in $\theta$ rotates $\Phi \rightarrow e^{i \theta} \Phi$\,, and $\bar u_1 \rightarrow e^{-i \theta} \bar u_1, ~\bar u_2 \rightarrow e^{i \theta} \bar u_2$\,. This symmetry is only broken by the off-diagonal entries of the Yukawa matrices, and as a result $\phi$ is a pNGB. 

Using the parameters justified above, the matrix \ref{eq:EffMassMat} is block diagonal, and only necessitates an $SO(2)$ rotation $O_{u}$ in the upper-left corner followed by a phase rotation. That is, one obtains an effective CKM matrix for the 3 SM generation that can be written as~\cite{Dine:2024bxv}:
\begin{equation}
	V = O_{u}^TP V^{d} \, ,
	\label{eq:EffCKM}
\end{equation}
where the matrix $P$ is simply
\begin{equation}
	P = \rm{diag}\left( 1, \exp\left(-i\ \left(\theta_{\rm CKM} + 2\frac{\phi}{f}
\right)\right),1 \right) \, .
	\label{eq:PMat}
\end{equation}
This means that, for a given $V^d$, the CKM matrix effectively has two free parameters left, $\theta_{\rm CKM}$ and the mixing angle in $O_u$:
\begin{equation}
	\theta_{12}^u = \frac{1}{2}\tan^{-1}(\frac{2f^2 g \tilde{g} m_c m_u}{f^2(g^2 m_c^2-\tilde{g}m_u^2 )+ 2\mu^2 (m_c^2-m_u^2)}) \, ,
	\label{<+label+>}
\end{equation}
where, in a slight abuse of notation, we denote with $g (\tilde{g})$ the only nonzero component of the vector $g(\tilde{g})$.

We write $V_d$ as 
\begin{equation}
	V^{d}=
\begin{pmatrix}
c_{12}^d c_{13}^d &
s_{12}^d c_{13}^d &
s_{13}^d  \\[6pt]
- s_{12}^d c_{23}^d - c_{12}^d s_{23}^d s_{13}^d  &
\;\;c_{12}^d c_{23}^d - s_{12}^d s_{23}^d s_{13}^d  &
s_{23}^d c_{13}^d \\[6pt]
s_{12}^d s_{23}^d - c_{12}^d c_{23}^d s_{13}^d  &
- c_{12}^d s_{23}^d - s_{12}^d c_{12}^d s_{13}^d  &
c_{23}^d c_{13}^d
\end{pmatrix}
\,,
\end {equation}
with $c_{12}^d= \cos \theta_{12}^d$, $s_{12}^d= \sin \theta_{12}^d$, and similarly for the other angles.
This is simply the standard SM parametrization, with the exception that we set $\delta =0$ since $V_d$ is by construction a real matrix.

It is interesting to show that even in the two generation limit, where the CKM phase is not physical, the Cabibbo angle depends on $\phi$. The limit can be obtained by taking $s_{13}^d=s_{23}^d=0$, leading to the following structure
\begin{align}
	|V_{us}|^2 &= \sin^2(\theta_{12}^u -\theta_{12}^d) + \sin^2\left(\frac{\theta_{\rm CKM}}{2} + \frac{\phi}{f}\right)\sin(2\theta_{12}^u)\sin(2\theta_{12}^d)\\
	|V_{ud}|^2 &= 1 - |V_{us}|^2 \,,
	\label{}
\end{align}
and the matrix is exactly symmetric.
Given the smallness of $\theta_{13},\theta_{23}$ in the SM, these are the leading contributions even in a realistic model.

Examining the expression for the mixing we would like to make the following points. First, if $\theta^{d,u}_{12} =n\pi/2$ for some integer $n$, \textit{i.e.} no mixing in either the up or down sector, $\phi$ does not enter the expressions for the CKM matrix elements. 
Moreover, $\theta_{\rm CKM} = 2 n \pi$ -- which implies an unbroken CP in the $\phi$ sector -- implies no time-independent shift from the SM expectation. 
In this case, the time-dependent piece would be of order $(\phi/f)^2$, which is expected because $\phi$ is a pseudoscalar and $|V_{ij}|^2$ is a CP-even quantity.
For nonzero $\theta_{\rm CKM}$, on the other hand, we obtain oscillations that are linear in $\phi/f$, because $\theta_{\rm CKM}$ is a CP-violating spurion.

For any value of the parameters, and ignoring the time dependence, the CP-violating phase can be eliminated as in the SM, since any $2\times2$ unitary matrix is real. 
However, in the presence of a time-dependent ULDM background, CP-violating observables can be defined even in a two generation limit, as was shown in ref.~\cite{Losada:2023zap} for the neutrino sector. The observables defined in that study can be adopted for the quark sector, but this is not the focus of this work.
\subsection{Oscillating CKM from Quadratically Coupled ULDM}
The above setup was motivated by the strong CP problem providing a construction in which either a pNGB or an ALP field plays the role of the CKM phase. It was interesting, however, to notice that the Cabibbo angle is a quadratic function of the corresponding angular variable, and therefore would oscillate with time, even in the two generation limit of the SM, where there is no physical time-independent CKM phase. 
However, when the field acquires a complex VEV, the coupling of the CKM matrix to the oscillating field becomes linear.
Such couplings are highly constrained, because at one higher loop they induce violation of equivalence principle (EP), or gravity-competing fifth forces~\cite{Dine:2024bxv}.
Moreover, they are bounded by \ce{^{229}Th} spectroscopy, as shown recently in ref.~\cite{Arakawa:2026mls}. 

Here, we construct another type of technically-natural model where the CKM elements depends quadratically on the pNGB ULDM field.
This class tries to avoid the EP and fifth-force bounds, by removing the long-range force mediated by a single exchange of $\phi$.
The idea is based on the fact that an ALP generically has quadratic coupling to CP-even SM scalar operators. 
These could naturally  arise from the presence of anomalous current, as in the case of the Peccei--Quinn symmetry~\cite{Peccei:1977hh}, yielding a quadratic coupling for the QCD axion~\cite{Kim:2022ype}, from other types of UV instantons (see e.g.~\cite{Holdom:1982ex,Choi:1988sy,Rubakov:1997vp,Gherghetta:2016fhp,Gaillard:2018xgk,Csaki:2019vte,Kivel:2022emq}), from variants of twin-symmetry~\cite{Delaunay:2025pho}, or from more general $Z_2$ symmetric structures~\cite{Banerjee:2022sqg}.
Indeed, for any scalar that enjoys a softly-broken shift symmetry the structure of this coupling is $(\partial\theta)^2\, O^{\rm SM}$, with $\theta=\phi/f$, and,  $O^{\rm SM}$ being a dimension-$d$, parity-even SM operator. 
By integration by part and field redefinition, this can be replaced with the dimension $d+2$ operator $\epsilon^2\theta^2\, O^{\rm SM}$,\footnote{As well as other higher dimension operators, such as $\theta^2 \Box O^{\textrm{SM}}$.} $\epsilon^2$ corresponding to a small parameter associated with the effective breaking of the shift symmetry. 
A lower bound of $\epsilon^2\gtrsim m^2/f^2$ is expected because the mass itself provides soft breaking of the shift symmetry~\cite{Banerjee:2022sqg}. 
However, a much bigger $\epsilon^2$ can be obtained by having further sources of symmetry breaking; for instance, for the celebrated QCD axion $\epsilon^2\sim m_{u,d}/\Lambda_{\rm QCD}$, which is larger by a factor $\sim (f/\Lambda_{\rm QCD})^4$.
Using the mechanisms discussed above, one can then have explicit shift-symmetry breaking while naturally keeping the mass low, as well as a quadratic coupling to the SM parametrically enhanced with respect to the linear.

Here we do not attempt to write a full UV complete model, and instead just add one such coupling, which leads to Cabibbo angle that depends quadratically on the ALP field.
Consider for instance a two generation Froggatt--Nielsen effective model, with the up and charm carrying different U(1) charges to account for the hierarchy of their masses and the smallness of the Cabibbo angle.  We add to the action an off-diagonal quadratic coupling $\epsilon^2 \theta^2 \tilde{H} \bar Q_1 c_R$, with $Q_1$ being an approximately first generation weak-doublet, $c_R$ an approximate charm singlet field, and $H$ the Higgs doublet. This induces the following form for the up-type mass matrix, in the basis where the down quark Yukawa is approximately diagonal,  
\begin{equation}
	Y_u \approx
\begin{pmatrix}
y_u & \lambda y_c \\
-\lambda y_u & y_c
\end{pmatrix} 
+
\begin{pmatrix}
0 & \epsilon^2 \theta^2 \\
0 & 0 
\end{pmatrix} \,.
\end{equation}
 Thus the mass matrix is not diagonalised by a rotation by $\lambda$, anymore, and as a result one measures an effective Cabibbo angle $\lambda_{\textrm{eff}} (\theta)$:
 \begin{equation}
	 \lambda_{\rm eff}
	 \approx 
	 \lambda \left(
		 1+ \frac{\epsilon^2 \theta^2}{\lambda y_c} 
		 \right) 
		 \rightarrow
	 \frac{\delta \lambda^{\max}}{\lambda}
	 \sim
 10^{3} \epsilon^2 \frac{\rho_{\rm DM}}{m^2 f^2}
	 \sim 
 10^{-3}\,\epsilon^2 \,\left(\frac{10^{-19}{\rm \, eV}}{m}\right)^2 \left(\frac{10^{10}{\rm \, GeV}}{f}\right)^2 \, ,
 \end{equation}
where we used that $\theta^2$ oscillates between $0$ and $\rho_{\rm DM}/(m f)^2$.

Several remarks are in order to contextualize these results.
The expression above are obtained at leading order in the Cabibbo-angle expansion, and subleading effects are expected to correct not only the off-diagonal but also the diagonal entries of the mass matrix.
As a consequence, quark masses should receive additional contributions suppressed by at least $\lambda \epsilon^2$.
In particular, the diagonal coupling $\lambda \epsilon^2 \theta^2 \tilde{H} \bar c_L c_R$ induces a gluon coupling when integrating out the charm~\cite{Shifman:1978zn,Kaplan:2000hh}:
\begin{equation}
	\frac{\beta_3}{2 g_3}\frac{2\epsilon^2 \lambda}{27 y_c}\theta^2 G G\,,
	\label{<+label+>}
\end{equation}
which induces stringent constraints. 

\noindent In light of the analysis presented below, together with the estimates of~\cite{Dine:2024bxv}, value of $\epsilon^2 \gtrsim 10^{-3}$ appear to be in reach of current flavor experiments.
This corresponds to an effective quadratic coupling that is significantly smaller than what is found to be natural in ref.~\cite{Delaunay:2025pho}, meaning that models without fine tuning can be probed at flavour factories.
Moreover, since the model only involves quadratic ULDM coupling, and the field oscillates around zero on account of having no vacuum expectation value, it is not subject to constraints from EP tests at tree level.
However, as we will see, EP bounds are reintroduced on account of the nontrivial field profile of $\phi$ on Earth~\cite{Hees:2018fpg}.

Having at disposal two different models that predict oscillations in the CKM, we now turn to observable consequences.
We will concentrate on the effect that oscillations of CKM have on the lifetime of mesons, specifically Kaons.
\section{CKM Oscillations and Meson Lifetimes}
\label{sec:Lifetime}
In this section, we collect some facts about the effect of varying $V_{\textrm{CKM}}$ on the lifetime $\tau$ of a particle that decays through a flavor-changing current, as well as the generic strategy to look for these effects.

\subsection{Time-Varying Lifetime}
A time-varying CKM element implies a time-varying effective lifetime:
\begin{equation}
	\tau(t_\textrm{obs}) \propto 
    |V_{\textrm{CKM}}(t_{\textrm{obs}})|^2 = |V_{\textrm{CKM}}(0)|^2 \left(1 + \delta \sin(f t_{\textrm{obs}})\right) \, ,
	\label{eq:EffLif}
\end{equation}
where $t_{\textrm{obs}}$ is the wall-clock time at which the CKM element is measured, and $f$ is the oscillation frequency, which is set by the model.
In particular, if the coupling between the CKM and the ULDM field is linear, the frequency is equal to the mass of the ULDM, $f^{\rm lin} = m_{\phi}$, while if the coupling is quadratic it is equal to twice the mass $f^{\rm quad} = 2 m_{\phi}$~\cite{Banerjee:2018xmn}.
Note that for consistency, we will need $\tau \ll 1/f$, such that the oscillations are essentially stationary during the $K$ decays.

In turn, the decay distribution of the particle is modified to be \begin{equation}
	p(t,t_{\textrm{obs}})  = \Gamma(t_{\textrm{obs}}) \exp(-t\Gamma(t_{\textrm{obs}} ))\, ,
	\label{eq:DecDistr}
\end{equation}

where $t$ is the proper time of the particle (distinct from the observation time $t_{\textrm{obs}}$), and $\Gamma \equiv 1/\tau$.

Consider the probability distribution in \cref{eq:DecDistr}, and let us for now disregard effects related to a finite observation time, which will be addressed in the next section.
For small $\delta$ we can expand in series to find the fractional difference with respect to a simple exponential decay:
\begin{equation}
	1 -
	\frac{p(t,t_{\textrm{obs}})}{p(t,0)} =
	\delta  (1-\Gamma t)
	\sin(f t_{\textrm{obs}}) 
	+ \mathcal{O}(\delta^2) \, .
	\label{eq:Sin}
\end{equation}
where $\Gamma \equiv \Gamma(t_{\rm obs}=0)$.

If, for experimental reasons, it is not possible to obtain data as a function of $t_{\textrm{obs}}$, and instead an average is performed, the effect is only visible at second order in $\delta$

\begin{equation}
	\int_0^{2\pi/f}\textrm{d}t_{\textrm{obs}}\, 
	\left(
	1 - 
	\frac{p(t,t_{\textrm{obs}})}{\Gamma \exp(-\Gamma t)} 
	\right)
=
	\delta^2 
	(\Gamma t -\frac{1}{2} - \frac{1}{4}\Gamma^2 t^2 )
+	\mathcal{O}(\delta^4) \, .
	\label{eq:RMS}
\end{equation}

\begin{figure}[htpb]
	\centering
	\begin{subfigure}[t]{0.49\textwidth}
		\centering
	\includegraphics[height=0.7\textwidth]{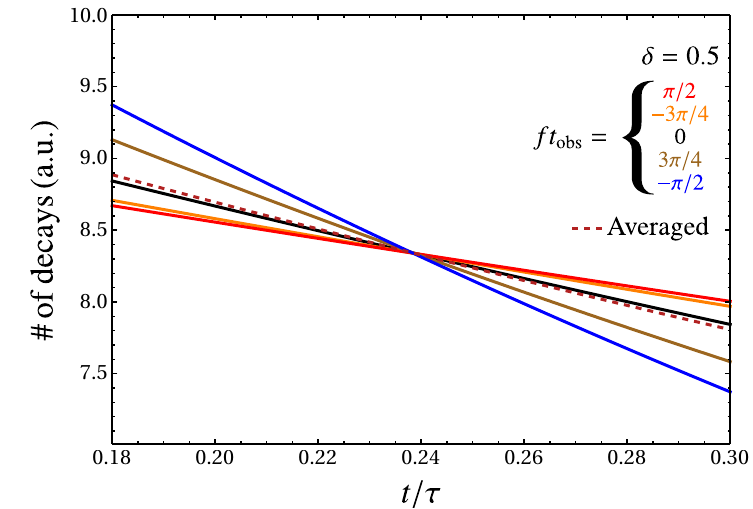}
	\end{subfigure}
	\begin{subfigure}[t]{0.49\textwidth}
		\centering
	\includegraphics[height=0.7\textwidth]{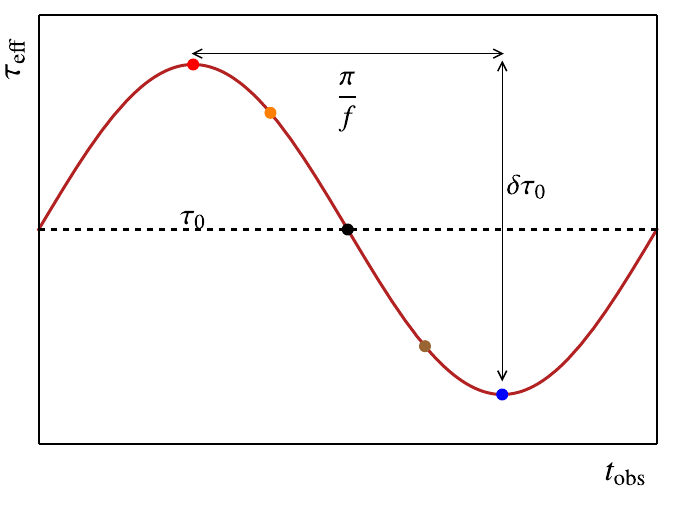}
	\end{subfigure}
	\caption{  \textbf{Left:} decay probability as a function of $t/\tau$ for fixed $\delta = 0.5$ and different $t_{\textrm{obs}}$ (continuous), as well as the average over one oscillation period (dashed). The black curve is also the SM expectation with time-independent $\tau$. The curves are all normalized to have the same integral in the observed region, corresponding to a situation where the total number of prepared particles is not known. 
\textbf{Right:} Effective lifetime as a function of $t_{\textrm{obs}}$.
    }
	\label{fig:DecayProb}
\end{figure}

We show in \cref{fig:DecayProb} how the decay probability looks at different $t_{\textrm{obs}}$, as well as the quadratic shift that happens when averaging. 
Interestingly, the average does not coincide with the SM expectation.
Nonetheless, as visible from  \cref{fig:DecayProb}, left-hand side, as well as comparing \cref{eq:Sin} and \cref{eq:RMS}, the best sensitivity to ULDM is gained from  a time-dependent analysis, which requires “snapshots” of the decay distributions to be taken.
We anticipate here an issue that will be crucial in the following section, namely that of the overall normalization and observation time.
Since in any realistic experiment decays can only be observed for an finite amount of time, not all prepared particles decay in the observation time. 
This is of course related to the lifetime under the assumption an exponential decay law, but as we will see in \cref{sec:Obs} knowing both can radically change the sensitivity of the experiment.

We now proceed by outlining the generic analysis procedure to detect or exclude the presence of $\phi$ interactions by looking at the time dependence of some observable that depends on $\tau_{\textrm{eff}}$. 

\subsection{Generic Measurement Strategy}
\label{sec:TimeSeries}
We seek to measure the time oscillation of a given observable $\mathcal{Q}(\tau;N_{\rm tot})$ that depends on the lifetime, as well as on the total number of Kaons involved in its determination, $N_{\rm tot}$.
We assume that $N_{\rm obs}$, the number of observed decays, is determined by some random process in time, which includes for instance time fluctuation of beam intensity and composition.
To first order in $\delta$, we find
\begin{equation}
	\mathcal{Q}(t) = \mathcal{Q}^{SM}(t,N_{\rm tot}) \left( 1 + C \delta \sin(f t) + \mathcal{O}(\delta^2)\right)\, ,
	\label{}
\end{equation}
with $C$ some constant depending on the specific observable.
In the ideal case, the best measurement strategy is to consider the Fourier transform of $\mathcal{Q}$:
\begin{equation}
	\mathcal{F} \left(\mathcal{Q}(t)\right)[\omega] =
	\mathcal{F}\left(\mathcal{Q}^{SM}\right)[\omega] 
	+ i \frac{C \delta}{2} \left(\mathcal{F}\left(\mathcal{Q}^{\rm SM}\right)[\omega - f] - \mathcal{F}\left(\mathcal{Q}^{SM}\right)[\omega + f]\right)
	+ \mathcal{O}(\delta^2) \, ,
	\label{}
\end{equation}
since multiplication by $\sin(f t)$ acts as a frequency shift by $\pm f$ in Fourier space.

We now make several assumptions, which will be relevant for the following sections. 
We assume that $\mathcal{Q}^{\textrm{SM}}$ has no intrinsic time dependence and that it is linear in $N_{\rm tot}$. 
Furthermore, we model $N_{\rm tot}$ as white noise with mean $\mu$ and non-constant part $n(t)$.
\begin{equation}
	\mathcal{Q}(t) = \mu\,n(t)\left(1 + \delta \cos(ft + \alpha)\right) \, ,
	\label{:q
	}
\end{equation}
where we will take $n(t)$ to be a gaussian white noise with variance $\sigma^2$, and $\alpha$ is an unknown phase.
We now define the finite-time Fourier transform of the mean-subtracted signal:
\begin{equation}
	X_T(\omega) = \frac{1}{T} \int_0^T \, {\rm d} t\, (x(t) - \mu) \exp(- i \omega t) \, ,
	\label{:}
\end{equation}
with $T$ the observation time. 
This is equivalent to the Fourier transform of the product of the signal times a square window function $\Theta(t)\Theta(T-t)/T$.
For convenience, let us write the finite-time Fourier transform of $1$:
\begin{equation}
	A_T(\omega) = 	\frac{1}{T} \int_0^T \, {\rm d} t\, \exp(- i \omega t) = \exp(- i/2 T \omega) \sinc(\omega T/2) \, ,
	\label{}
\end{equation}
where $\sinc(x) = \sin(x)/x$.
The finite-time Fourier transform of $x(t) - \mu$ can be expressed as the convolution between $A_T(\omega)$ and the infinite-time Fourier transform of $x(t)-\mu$.

We will be interested in the power spectral density $S(\omega)$, defined as the expectation value of $|X(\omega)|^2$.
If $\delta =0$, the power spectral density of the background is by definition $\sigma^2$
\begin{equation}
	S_{\rm bkg} = \sigma^2 \, .
	\label{}
\end{equation}
On the other hand, the signal has power spectral density is
\begin{equation}
	S_{\rm sig} = \frac{\delta^2}{4} (|A_T(f- \omega )|^2 + |A_T(f+ \omega)|^2 + 2\Re(\exp(2 i \alpha) A_T(f- \omega) \bar{A}_T(f+ \omega)) \, .
	\label{<+label+>}
\end{equation}
If we only consider positive frequencies, only the terms involving $\sinc(f-\omega)$ are relevant, and thus we lose any dependence on $\alpha$:
\begin{equation}
	S_{\rm tot} = S_{\rm bkg} + S_{\rm sig} \approx \sigma^2 + \frac{\delta^2}{4}\sinc^2\left(T(f-\omega)/2\right) \, .
	\label{<+label+>}
\end{equation}
We note that the interference between the background and the signal is zero on account of having subtracted the background mean.
We will use the power spectral density to obtain exclusion limits on $\delta$.
To do so, we construct the test statistic
\begin{equation}
	Z \equiv \frac{2 |\mathcal{E}(X(\omega))|^2}{\sigma^2} \, ,
	\label{<+label+>}
\end{equation}
which follows a $\chi^2$ distribution with 2 degrees of freedom (real and imaginary part of $X_T$) in the background-only hypothesis. 
The signal, on the other hand, is a deterministic spike in the power spectrum.
For confidence level $68\%(95\%)$ we then require
\begin{equation}
	\frac{2 \delta^2 }{\sigma^2}\sinc\left( T(\omega-f)/2 \right) > 2.28(6) \, ,
	\label{eq:z}
\end{equation}
that is, the smallest $\delta$ that can be excluded at $68\%(95\%)$ confidence level is equal to $1.07(1.73)\sigma$ if the Fourier transform is performed at $\omega \approx f$.
Note that, since $f$ is unknown, we need to scan over different values $f\in[f_{\min},f_{\max}]$.
The confidence levels obtained using this procedure is thus to be interpreted as local.
The value of $f_{\min}$ is determined by the running time of the experiment, since at least one full oscillation should occur to observe the effect.
As for $f_{\max}$, the Nyquist frequency sets a higher limit for equally-spaced observations, but for irregularly-spaced observations one can push $f_{\max}$ to much higher values given by the inverse of the timing uncertainty of the observations~\cite{Eyer:1998jr,KoenLS}.
In any case, given how mesons are produced at accelerator-based experiments (see \cref{sec:NA62}), we expect $f_{max} \lesssim 1/t_{\rm spill}$, the inverse of the proton spill, or whichever statistical process produces the mesons.
The correct tool to obtain the power spectrum of irregularly-spaced data is the Lomb--Scargle periodogram~\cite{Lomb:1976wy,Scargle:1982bw}, whose reach in $\delta$ can still be approximated as above.\footnote{Other techniques such as the wavelet transform or empirical mode decomposition~\cite{Huang:1998emd} can also be employed.}

If the signal in the time-domain is measured $N$ times independently with an uncertainty $\sigma_{\textrm{one shot}}$, then we obtain $\delta_{\min} = 1.07 (1.73) \sigma_{\textrm{one shot}}/\sqrt N$.
As we will explain in detail in \cref{sec:NA62}, the nature of the experiment and our observables impose that we collect a certain number of event.
That is, a single observation is obtained by grouping a certain number of $K$ decays.
The scaling of $\delta_{\min}$ is then reassuring, because it does not depend on how the grouping is performed if $\sigma_{\textrm{one shot}} \propto 1/\sqrt{N_{\rm obs}}$.
Indeed if the data is re-grouped such that a single observation is obtained out of $c N_{\rm obs}$ events, then $\sigma_{\textrm{one shot}}$ increases by $\sqrt{c}$, but $N_{\rm obs}$ decreases by $c$, hence $\delta_{\min}$ stays the same.
If this is not the case, the analysis becomes more involved, and an optimum binning needs to be determined. 
This assumption is trivially broken if the expected number of events is of order one, since empty bins have an undefined uncertainty.
Similarly, we will see in \cref{sec:Obs} that the lifetime does not necessarily respect this assumption either.

We expect systematic uncertainties in the determination of $\mathcal{Q}$ to have a relatively low impact on the final sensitivity. 
Consider, for instance, that a systematic effect shifts the observable up or down with respect to its true value. 
This shift would presumably be constant over the whole data-taking period and thus would not affect a cosine fit, since it only changes its mean value around which the oscillations are supposed to take place.
Similarly, if systematic effects such as acceptance cause the decay distribution to deviate from a simple exponential, the process can be adapted to look for variation in the empirically determined distribution, instead of the theoretical one.
Even in the absence of control over the systematic effects, a temporal variation in the true distribution of decays should be reflected in the empirical one.
A non-observation of the latter can then be translated into a limit on the former.

If it is not possible to measure the time series of $\mathcal{Q}$, one can average over a full period.
Perhaps counter-intuitively, the deviation does not vanish, as can be seen by performing the expansion up to at least quadratic order, see \cref{eq:RMS}.
Then, the whole data set deviates from the exponential decay law only quadratically in $\delta$, and thus one has reduced sensitivity, by comparing the residuals of a simple exponential fits with \cref{eq:RMS}. 
The sensitivity might be further limited by systematic uncertainty, due to \textit{e.g.}~not-constant acceptance in proper time, which might be impossible to distinguish from the effect of a quadratic shift.

We now describe the observable we propose for the search.
As hinted above, the lifetime seems to be the most apt observable.
However, an important feature of flavor experiments is that the mesons are boosted with respect to the laboratory frame, and as a result not all particles decay inside the region where they can be detected.
We will see in the next section that this greatly impact the choice of an observable with lowest $\sigma$.

\section{Choice of Observable}
\label{sec:Obs}
In ref.~\cite{Dine:2024bxv}, the authors advance the proposal to measure the time variation of the lifetime of the $K^+$, $\tau_K$, at the SPS experiment NA62 as a probe for ULDM models that couple linearly to CKM matrix elements.
We postpone more details about the experiment until \cref{sec:NA62}, but for now we note that because of their large boost, only $\sim 9\%$ of the $K^+$ decay within the fiducial volume of the experiment.
It was already noted in ref.~\cite{Dine:2024bxv} that this might worsen the reach of the experiment, and in \cref{sec:Unc} we show that indeed this is the case: the uncertainty on $\tau_K$ scales with the number of $K$ as of $C/\sqrt{N_{\rm obs}}$, where $C\sim 30$ for NA62.
As a result, the reach computed in ref.~\cite{Dine:2024bxv} overestimates the sensitivity by a factor of $15$.

This worsening of the uncertainty can be understood as a consequence of the degeneracy between the overall normalization (\textit{i.e.}~the total number of particles produced) and the lifetime. 
As a result, knowledge of the total number of produced $K$ can bring the uncertainty on the lifetime back to the usual $1/\sqrt{N_{K}}$ estimate.
However, at an experiment such as NA62 the total flux cannot be known precisely: while any $K$ that flies in the fiducial volume is, in principle, seen by the detector, events where the $K$ does not decay are not registered on tape by virtue of a trigger requirement condition.\footnote{Note that random and periodic triggers are introduced for calibration and monitoring purposes, but not useful for our purposes.}
In \cref{sec:Deg} we study the impact of the uncertainty on the flux, treating it as a nuisance parameter, and conclude that the interplay between this and the statistical uncertainty makes the lifetime a suboptimal observable to detect the effects of time-varying CKM matrix elements.
In \cref{sec:Counting}, we then consider the number of $K$ decay in the decay volume as a proxy for the CKM, finding it to be a suitable observable for our purposes.

\subsection{Statistical Uncertainty on Lifetime}
\label{sec:Unc}
The statistical uncertainty $\sigma_\Gamma$ on the parameter $\Gamma = 1/\tau$ is obtained by fitting a set of decay times $\{t_i\}_i$ to an exponential distribution and computing the Fisher information of the resulting maximum likelihood estimate (MLE)~\cite{CensoredTruncated}.
We review MLE in \cref{sec:MLE}.

Since particle decays can only be observed if they happen during a finite window, we impose for any $i$ that $t_i$ is within the interval $[T_1,T_2]$ of length $\Delta T \equiv T_2 - T_1$. 
As a result, the probability to observe a decay at time $t=t_i$ is the conditional probability
\footnote{$\tilde{p}(t)$ has dimension of inverse time, since strictly speaking it is a probability density. An actual probability can be be obtained by integrating over some time interval. We will be somewhat cavalier about this while computing the logarithm of this quantity, but of course one can have dimensionless quantities by expressing everything in unit of a reference time.}
\begin{equation}
\tilde{p}(t_i|\Gamma,T_1,T_2)) = \frac{p(t_i|\Gamma, T_1=0,T_2=\infty)}{p(T_1< t_i <T_2|\Gamma)} = \frac{\Gamma \exp(-\Gamma t_i)}{\exp(-y_1)-\exp(-y_2)} \,,
\end{equation}
where we defined $y_{1/2} \equiv \Gamma T_{1/2}$.
At NA62, considering the boost factor $\gamma \approx 150$, and the experimental geometry (which we will review in \cref{sec:NA62}), we have $y_{1}\approx 0.18, y_{2}\approx 0.3$.

The log-likelihood of a dataset $\{t_i\}_i$ of $N_{\rm obs}$ observed decays is:
\begin{equation}
    \ln L(\{t_i\}_i,N_{\rm obs}|\Gamma,y_1,y_2) = N_{\rm obs} \left(\ln\Gamma - \Gamma t^* - \ln(\exp(-y_1)-\exp(-y_2))\right)\, ,
 \label{eq:Truncated}
\end{equation}
where $t_* = \sum_i^{N_{\rm obs}} t_i/N_{\rm obs}$ is the mean of the observed decay times.
Using the definition of the MLE expected uncertainty, \cref{eq:MLEUnc}, the relative uncertainty is~\cite{CensoredTruncated,KLOE:2007wlh}:
\begin{equation}
	\frac{\sigma_\Gamma}{\Gamma} =
	\sqrt{ \frac{1}{N_{\rm obs}} } 
	\frac{\sinhc(\frac{\Delta y}{2})}{\sqrt{\sinhc(\frac{\Delta y}{2})^2-1}} \, ,
	\label{eq:RelUncLif}
\end{equation}
where $\sinhc(x) = \sinh(x)/x$ and $\Delta y = y_2 - y_1$. 
The relative uncertainty on $\tau$ is the same, since $\tau = 1/\Gamma$.
Importantly, the uncertainty only depends on $\Delta y$ and not on $y_1 + y_2$.

We can approximate $\sigma_\Gamma$ as follows:

\begin{equation}
    \frac{\sigma_\Gamma}{\Gamma} = \frac{1}{\sqrt{N_{\rm obs}}}\times
    \begin{cases}
        \frac{2\sqrt 3}{\Delta y} \quad & \textrm{if}\quad \Delta y \lesssim 4 \\
        1 \quad & \textrm{if}\quad \Delta y \gtrsim 4
    \end{cases} \quad .
    \label{eq:approxSigma}
\end{equation}

This means that when comparing two experiments with different $N_{\rm obs}$ and  $\Delta y$, the usual $1/\sqrt{N_{\rm obs}}$ scaling can be applied only if both experiments have large $\Delta y$, and otherwise \cref{eq:RelUncLif,eq:approxSigma} need to be used to estimate the sensitivity.
Using the NA62 values $y_{1} \approx 0.18,\, y_{2}\approx 0.3$, we see that there is a factor of 30 with respect to the naive $1/\sqrt{N_{\rm obs}}$ estimate.
We show in \cref{sec:Intuitive} that the ``penalty'' factor  when $\Delta y$ is small can be understood as a consequence of degeneracy between $\tau$ and $N_{\rm tot}$, the total number of particles produced, including those that decay before $T_1$ or after $T_2$. 

In the next section we show how knowledge of $N_{\rm tot}$ restores the usual $1/\sqrt{N_{\rm obs}}$ scaling.

\subsection{Breaking the Degeneracy}
\label{sec:Deg}
If not only the time of decays, but also the total number of prepared particles, $N_{\rm tot}$ is known, we are in what is known as a “censored” case, as opposed to the “truncated” case in which $N_{\rm tot}$ is not known~\cite{CensoredTruncated}. 
For censored dataset, the correct scaling for the uncertainty is the naive $1/{\sqrt{N_{\rm obs}}}$.
Intuitively, this is because  knowledge of the total number of particles fixes the tail of the distribution, allowing to recover the usual scaling.  

Explicitly, the likelihood obtains contribution from $N_{\rm obs}$ observed decays, each contributing $p(t_i|\Gamma)$ (note the absence of tilde), and from $N_{\rm tot}-N_{\rm obs}$ unobserved decays, which each contribute $1-\int_{T_1}^{T_2} p(t|\Gamma) {\rm d} t$:
\begin{equation}
	\ln{L(\{t_i\}_i,N_{\rm obs},N_{\rm tot}|\Gamma, T_1,T_2)} = N_{\rm obs}(\ln(\Gamma) - \Gamma t^*) + (N_{\rm tot}- N_{\rm obs})\ln(1- \exp(-y_1) + \exp(-y_2))  \, .
	\label{eq:Censored}
\end{equation}
The Fisher information is:
\begin{align}
	\mathcal{I} &=\mathds{E}[\sqrt{N_{\rm obs} \Gamma^2}] \, .
	\label{<+label+>}
\end{align}
Note that, in principle $\mathds{E}[{N_{\rm obs}}] = N_{\rm tot}(\exp(-y_1)- \exp(-y_2)) \neq N_{\rm obs}$, but if $N_{\rm tot}$ is well measured and $N_{\rm tot}, N_{\rm obs} \gg 1$, the two values do not differ by much, so we recover the usual $1/\sqrt{N_{\rm obs}}$ scaling for the relative uncertainty~\cite{CensoredTruncated}.
However, $N_{\rm tot}$ cannot be known exactly at experiments because an event is registered on tape only if a decay is detected in the fiducial region. In \cref{sec:UncNK} we study how imperfect knowledge of $N_{\rm tot}$ affects the uncertainty estimate.

\subsection{Imperfect $N_{\rm tot}$ Knowledge}
\label{sec:UncNK}
In this subsection, we show how to account for imperfect knowledge of $N_{\rm tot}$, by explicitly considering it as a nuisance parameter.
We introduce a probability distribution for $N_{\rm tot}$, $p\left( N_{\rm tot} \right)$.
The case of perfectly known $N_{\rm tot}$, corresponds of course to $p(N_{\rm tot}) = \delta_{N_{\rm tot}}^{N_{\rm tot}^*}$.
If the normalization $N_{\rm tot}$ is measured independently, its uncertainty can often be treated as Gaussian. We write
\begin{equation}
L(\Gamma, N_{\rm tot}) = p(N_{\rm tot}) \cdot L(\{t_i\}, N_{\rm obs}, N_{\rm tot} \mid \Gamma, T)\, ,
\label{eq:likelihoodGauss}
\end{equation}
then, one finds the MLE estimate by maximizing over both $\Gamma$ and $N_{\rm tot}$.
The uncertainty is found by the use of the profile likelihood~\cite{PDGStat}, but one can simplify the profiling procedure in the quadratic approximation, in which case the profiled uncertainty can be taken to be (see \cref{sec:MLE}):
\begin{equation}
I_{\Gamma,\mathrm{Fisher}} = -\frac{\partial^2 \ln L}{\partial \Gamma^2} + 
\frac{\left(\partial^2 \ln L / \partial \Gamma \partial N_{\rm tot} \right)^2}{\partial^2 \ln L / \partial N_{\rm tot}^2}\, .
\label{eq:FisherProfile}
\end{equation}
For ease of understanding, we can set $y_1 =0$ for now, as well as using the expected value $N_{\rm tot} = N_{\rm obs}/(1-\exp(-y_2))$.
This yields a relative uncertainty equal to
\begin{equation}
	\frac{\sigma_{\Gamma}}{\Gamma} =
	\left(
	N_{\rm obs}
-
\frac{y_2^2 N_{\rm obs}^2 \sigma_N^2}{
	N_{\rm obs}^2 \sigma_N^2
	\left( 
		\psi^{(1)}(1 + \frac{N_{\rm obs}}{\exp(y_2)-1}) -
		\psi^{(1)}(1 + \frac{N_{\rm obs}}{\exp(y_2)-1} + N_{\rm obs})
	\right)
	+ 
	(\exp(y_2) -1)^2
	}
\right)^{-\frac{1}{2}},
	\label{eq:RelUncLifGauss}
\end{equation}
where $\psi^{(1)}(x) = \mathrm{d}^2\Gamma(x)/\mathrm{d}x^2$ is the trigamma function, and we set $\sigma_N = \sigma/N_{\rm tot}$.
It is easy to see that for small $N_{\rm obs} \sigma_N$ we get
\begin{equation}
	\frac{\sigma_{\Gamma}}{\Gamma} =
	\left(
		N_{\rm obs} - \frac{N^2_{\rm obs}\sigma_N^2 y_2^2}{(\exp(-y_2)-1)^2} + \mathcal{O}(N_{\rm obs}^4\sigma_N^4)
\right)^{-\frac{1}{2}},
	\label{<+label+>}
\end{equation}
which simply tracks $1/\sqrt{N_{\rm obs}}$.
When $N_{\rm obs}\sim\sigma_N^{-2}$, the two contributions in the above expression balance each other approximately, and indeed a nearly-constant regime can be found starting from
$N_{1}\sim \exp(-y_2)/\sigma_N^2$, where $\sigma_\Gamma/\Gamma \sim y_2^2/(12\sigma_N^2)$, independent of $N_{\rm obs}$ as long as $y_2\ll 1$.
At large $N_{\rm obs}$, on the other hand, the uncertainty matches \cref{eq:RelUncLif}, recovering the same uncertainty as in the unknown $N_{\rm tot}$ case.
The turnover can be computed by imposing that the nearly-flat and $\sqrt{12/(y N)}$ approximate behavior match.
Namely, the near-plateau ends when $N_{\rm obs}$ is equal to
\begin{equation}
	N_2 \sim \frac{12}{y_2^2\sigma_N^2}(1-y_2) + \mathcal{O}(y_2^0) \, .
	\label{<+label+>}
\end{equation}

It is important to note that, in the plateau region, the assumption $\sigma_{\textrm{one shot}} \propto 1/\sqrt{N_{\rm obs}}$ is not true, hence $\delta_{min}$ is not invariant under data re-grouping, see the discussion under \cref{eq:z}.
In the asymptotic limit of large statistics -- which is a prerequisite for the MLE to be an unbiased estimator -- we always obtain an uncertainty that is much larger than the naive $1/\sqrt{N_{\rm obs}}$.

\begin{figure}[htpb]
	\centering
	\includegraphics{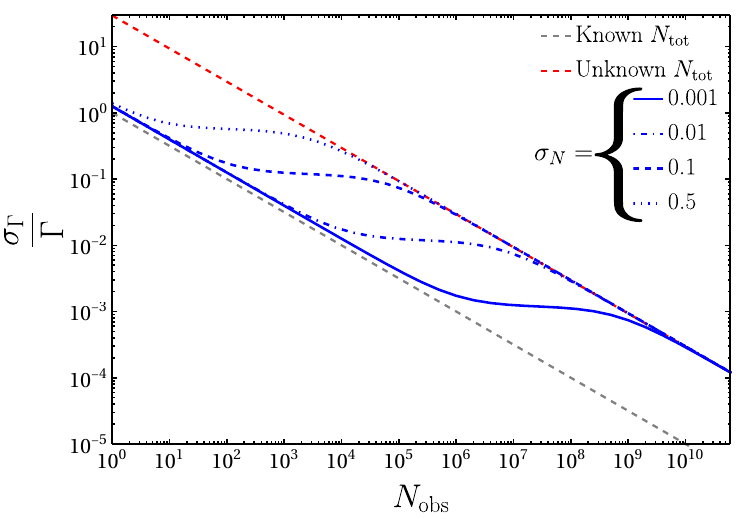}
	\caption{Relative uncertainty as a function of the total number of particles, fixing $y_{1,2}$ to the NA62 values. The dashed red line is the uncertainty obtained using a truncated exponential, equivalent to having no knowledge of the total number of particles, while the dashed gray line is the uncertainty in the case of perfectly known total number of particles. The blue lines are obtained using a gaussian probability distribution for $N_{\rm tot}$, with different choices of $\sigma_N$, see main text.}
	\label{fig:Unc}
\end{figure}

We have checked that these results are not an artifact of the approximations used by computing both the proper profiled likelihood uncertainty, and the Bayesian marginalisation, see \cref{app:LifeUnc}, which agree both qualitatively and quantitatively.
We show the uncertainty, using the NA62 parameters for $y_{1,2}$ in \cref{fig:Unc}.
We should note that, since $y_1 \neq 0$, the small $N_{\rm obs}$ trend is actually $c(y_1,y_2)/\sqrt{N_{\rm obs}}$, where
\begin{equation}
	c(y_{1},y_{2}) \approx {1+ y_1 + y_1 y_2} + \mathcal{O}(y_1^2) \, ,
	\label{<+label+>}
\end{equation}
which is strictly larger than 1, but only by a small amount $\propto y_1$.
This is the only quantitative difference with the $y_1=0$ case.
Indeed, at large $N_{\rm obs}$, the dependence on $y_1+y_2$ vanishes, and one obtains~\cref{eq:RelUncLif} even when considering $y_1\neq0$.

It should be noted that the computations in this section are, strictly speaking, only valid asymptotically.
The small- and intermediate $N_{\rm obs}$ regime should be treated with more care~\cite{Cowan:2010js}.
In any case, we can safely conclude that the lifetime is not suited for a search of oscillations of CKM elements at experiments with a small $\Delta y$ such as NA62.
We then turn to another, more suitable observable.

\subsection{Oscillations in the Counting}
\label{sec:Counting}
Since varying the lifetime affects the number of particles that decay in the window $[T_1,T_2]$, $N_{\rm obs}$ oscillates as a function of $t_{\rm obs}$.
We do not use this information to extract the effective lifetime $\tau(t_{\textrm{obs}})$, but instead directly look for oscillations in the count.
This procedure has the advantage of not needing to know $N_{\rm tot}$, but one has to assume that it is constant, or at the very least that systematic effects do not cause it to oscillate at frequencies $\sim f$, see the discussion at the end of \cref{sec:TimeSeries}.
It should also be noted that the production mechanism of $K$ at NA62 is a pure QCD process, so the oscillations in the CKM elements will not affect it.
Perhaps surprisingly, we show that the sensitivity to CKM oscillations is comparable to the naive $1/\sqrt{N_{\rm obs}}$.

To estimate the sensitivity, we first compute the shift ($\Delta N_{\rm obs}$) as a result of the effective substitution $\tau \to \tau \left(1+\delta\cos(ft)\right)$:
\begin{align}
\nonumber  
	N_{\rm obs} \to &N_{\rm obs}\left( 1 + \delta  \dfrac{y_2 - y_1 \exp(\Delta y)}{1-\exp\left(\Delta y\right)} \cos(ft) \right)+ \mathcal{O}\left(\delta^2\right) \, .
	\label{eq:NobsSens}
\end{align}
The sensitivity to $\delta$ scales simply as $1/(\sqrt{N}\Delta N)$.
For NA62, we then obtain 
\begin{equation}
	\delta^{\max}\approx \frac{1.3}{\sqrt{N_{\rm obs}}}\, .
	\label{<+label+>}
\end{equation}
combined with \cref{eq:z}, this means that the 68\%(95 \%) CL uncertainty is $1.4 (2.25)/\sqrt{N_{\rm obs}}$, broadly in agreement with the naive estimate of~\cite{Dine:2024bxv}. 

It is interesting to note that, for $y_1 < 1$, there are values of $y_2$ for which the shift in $N_{\rm obs}$ drops to 0.
This is because it is given by a competition between the shift in the expected number of survivors at $T_1$ and at $T_2$, and for some combinations of $y_1,\,y_2$ one gets cancellation between the two.
This in turn would imply that certain experiments might be insensitive to the oscillation in the CKM\@.
Interestingly, this could to distinguish a claimed sign from unknown systematics, by artificially changing $y_{1},y_{2}$ within the data analysis procedure.

At NA62, another appealing possibility is to measure $N_{\rm CEDAR}$, the number of $K$ mesons that pass by the first detector performing particle identification (PID), see \cref{sec:NA62}.
We define $y_{\rm CEDAR}\approx 0.13$ in analogy with $y_{1},y_{2}$.
The same computation as above, with $y_{1} \to 0, y_{2}\to y_{\rm CEDAR}$, gives
\begin{align}
	N_{\rm CEDAR} &\to N_{\rm CEDAR} \left(1 -  \delta \frac{y_{\rm CEDAR} }{1-\exp\left(y_{\rm CEDAR} \right)} \cos (ft)+ \mathcal{O}\left(\delta^2 y_{\rm CEDAR}^2)\right)\right)\\
	\delta^{\max}&\approx\frac{7.9}{\sqrt{N_{\rm CEDAR}}} \approx \frac{2.6}{\sqrt{N_{\rm obs}}}\,,
	\label{eq:NCEDARSens}
\end{align}
which is a factor of $2$ worse than the above, but might be used as a cross-check of systematic uncertainty, moreover $N_{\rm CEDAR}$ could be an easier quantity to measure than $N_{\rm obs}$, since it depends on only one detector.

Having singled out Kaon counting as a particularly sensitive and apt observable, before concluding we collect some practical information about the NA62 experiment, to better explain the challenges of this measurement.
\section{The Kaon factory NA62 and `Kaon counting'}
\label{sec:NA62}
NA62 is a fixed-target experiment located in CERN's north area which makes use of the SPS accelerator. 
It was built to measure precisely the branching ratio ${\cal B}(K^+\to\pi^+\nu\bar\nu$).

To measure this small branching ratio, the experiment makes use of a \SI{75}{GeV} unseparated Kaon beam whose decays are recorded over an effective decay region $L\approx \SI{65}{m}$, equipped with tracking, vetos and calorimetry, see ref.~\cite{NA62:2017rwk} for details.

	\begin{figure}[htpb]
		\centering
	\begin{tikzpicture}[scale=1.2,
    lbl/.style={font=\small}]

  \draw[->,thick]
    (-0.2,0)--(9.5,0)
    node[right,font=\small]{$z$};

  \fill[black](-0.1,-0.1) rectangle (0.1,0.1);
  \node[lbl,below] at (0.0035,-0.55){target};

  \fill[myred!35](0.33,-0.5) rectangle (1.00,0.5);
  \node[lbl,above] at (0.67,0.55){achromat};

  \fill[myyellow!80](2.13,-0.5) rectangle (2.27,0.5);
  \node[lbl,above] at (2.20,0.55){CEDAR};

  \fill[mygreen!30](3.40,-0.5) rectangle (5.57,0.5);
  \draw[mygreen,very thick]
    (3.40,-0.5) rectangle (5.57,0.5);

  \node[lbl,above] at (4.49,0.55)
    {fiducial decay volume};

  \fill[gray!45](8.67,-0.5) rectangle (9.00,0.5);
  \node[lbl,above] at (8.835,0.55){XION2};

  \draw[myblue,<->,thick]
    (0,1.1)--(2.20,1.1)
    node[midway,above,font=\small]
    {$D=cT_{\rm CEDAR}$};

  \draw[mygreen,<->,thick]
    (3.40,-1.)--(5.57,-1.)
    node[midway,below,font=\small]
    {$L=c\,\Delta T$};

\end{tikzpicture}\caption{Schematics  of the relevant elements of the NA62 experiment.
        The target for Kaon production is located at $z=\SI{0}{m}$. An achromat selects a \SI{75}{GeV} beam. Approximately $D\approx\SI{70}{m}$ behind the target, a differential Cherenkov with achromatic ring focus (CEDAR), with its photon detection system (KTAG) is used for Kaon identification, shown in purple in the above drawing.
		Approximately \SI{102}{m} behind the target, the fiducial decay volume begins.
        The black arrows indicate travel distances of Kaon to its identification $D$, and typical distances $L$  over which decay points can be selected.
	}
		\label{fig:detectorNA62}
	\end{figure}
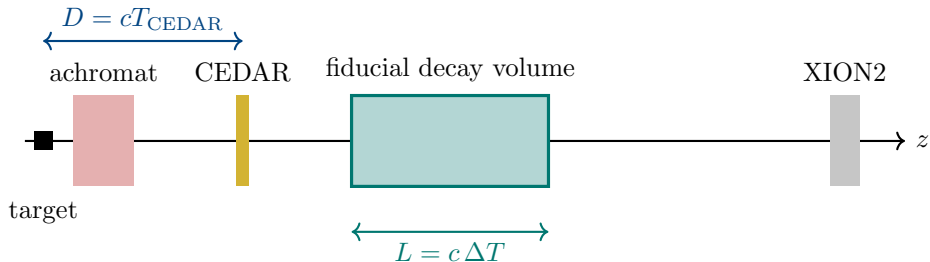

A detector schematic is 
shown in \cref{fig:detectorNA62}.
The unseparated Kaon beam of NA62 is produced approximately $\sim\SI{102}{m}$ upstream of the experiment's decay volume:
The \SI{400}{GeV} SPS proton beam is directed on to a beryllium target (labeled as `target' and located at \SI{0}{m} in the coordinate system of the experiment, see left part of \cref{fig:detectorNA62}).
In the interaction of the proton beam with the target, a spectrum of mesons is produced. 
In~\cite{Atherton:1980vj}, such a spectrum has been characterized for an equivalent configuration.

After the target, it follows a set of magnets and collimators act as achromats:
A beam of particles with a total momentum of $\sim \SI{75}{GeV}$ is selected and guided to the decay region.
The ‘nominal’ proton rate on T10 is $3\times 10^{12}$ protons per pulse over a \SI{4.8}{s} of `flat top' (dubbed `spill').

The main components of the  resulting secondary beam are $\pi^+$, $p^+$ and $K^+$. Due to decays, the relative composition of the beam is location-dependent: the fraction of $K^+$ in the beam is a bit over 6\% at the production point and a little less than 5\% at the exit of the decay volume. 
The particle identification (PID) is performed through a differential Cherenkov with achromatic ring focus (CEDAR) 
(purple in \cref{fig:detectorNA62}), see also \cite{NA62:2023mud}.
The CEDAR detector is combined
with a purpose-built photon detection system (KTAG) located around \SI{70}{m} downstream of the target.
The KTAG provides a precise kaon time reference for event reconstruction. The system is required to identify $K^{+}$  with efficiency above 95\%, and the kaon–pion separation must be better than $10^{4}$.

Whereas NA62's purpose normally is to measure precisely the decay of the positive Kaons, by contrast, here we are interested in the \textit{non-decayed} beam particles that traverse the CEDAR\@. A measure of the \textit{non-decayed} particles at the end of the beam-line is the so-called XION2 counter at around \SI{265}{m} behind the target, which, however has no PID capabilities.

For the purpose of this paper, we are less interested in the Kaon distribution or its fluctuation along the duration of the spill, but rather on possible variations from one spill to another, or even longer time periods. For this, fluctuations on the Kaon distribution primarily depend on the possible fluctuations on the initial proton beam.

A dominant source of such fluctuations are the variations of the incoming proton intensity on the target.
When the SPS accelerator is running in stable conditions, the proton intensity can be estimated by a gaussian over limited periods. In practice however,
there are slow drifts over hours and days and also after super-cycle changes~\cite{Lau}.
Such fluctuations on proton intensity will be the main factor determining the fluctuation of Kaons produced in the target, the production being a statistical process.
Thus, we expect the relative fluctuations in the Kaon distribution, normalized to the number of protons, to follow $\sim 1/\sqrt{N_{p}}$.

In reality, for a ULDM search as the one we advocate for in \cref{sec:Obs}, one will have to rely on data:
For any event triggered in the experiment, the number of Kaons present in the trigger window (beyond the decaying Kaon that caused the trigger) can be counted. 
Such studies are already on the way.

\section{Results}
\begin{figure}[htpb]
	\centering
	\includegraphics[width=\textwidth]{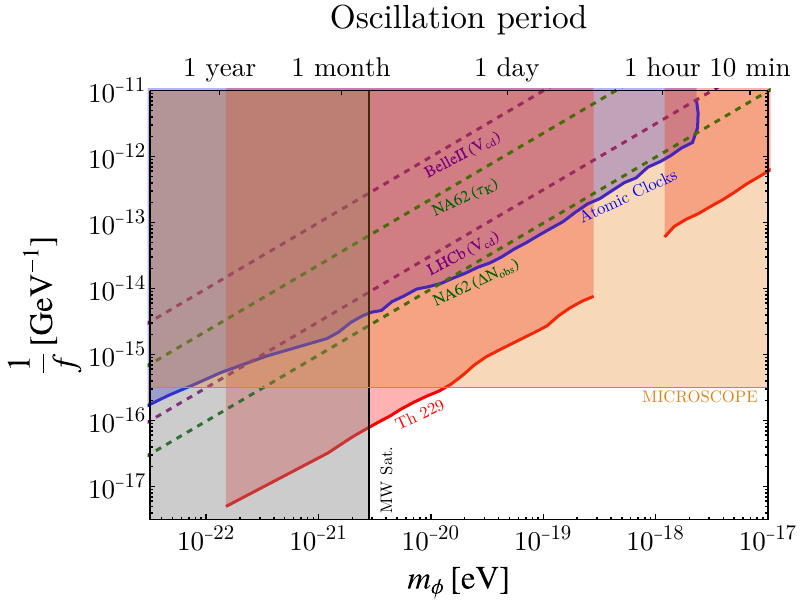}
	\caption{Bounds for the linear model. We show in green the estimated sensitivity to $V_{us}$ at NA62 (with $N_{\rm obs }= 10^{13}$) using both the lifetime $\tau_K$ and the number of observed $K$, $N_{\rm obs}$, while in purple we show the estimated sensitivity of LHCb (with $10^{13}$ D-meson decays) and BelleII (with $10^{10}$ decays) to oscillations of $V_{cd}$. The red and blue lines are the limits coming from atomic clocks and the \ce{^{229}Th} isomeric transition, respectively~\cite{VanTilburg:2015oza,Hees:2016gop,Kennedy:2020bac,BACON:2020ubh,Sherrill:2023zah,Arakawa:2026mls}. The gray shaded region is the constraint from the observation of Milky Way satellite galaxies~\cite{DES:2020fxi}, while the yellow shaded region is the EP bound~\cite{MICROSCOPE:2022doy}.}
	\label{fig:LinBounds}
\end{figure}

\begin{figure}[htpb]
	\includegraphics[width=\textwidth]{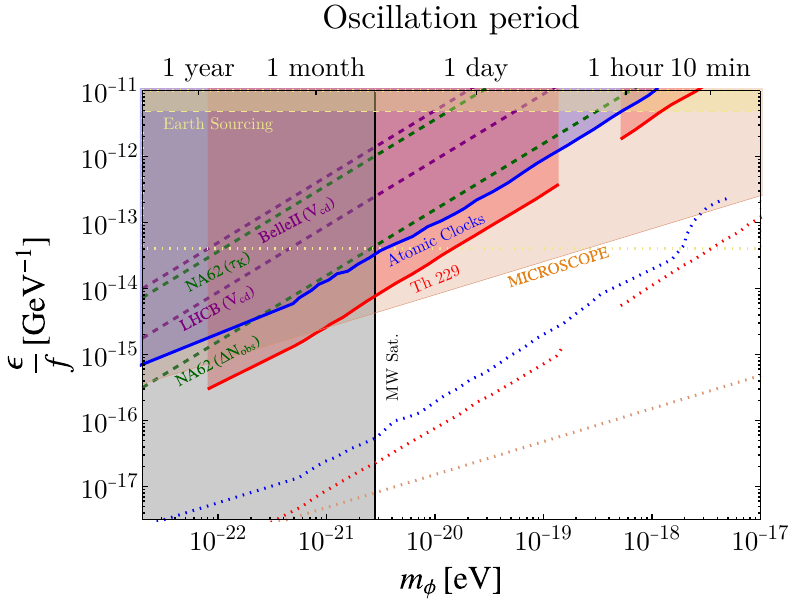}
	\caption{Same as \cref{fig:LinBounds}, but for the quadratically-coupled model. Solid lines constraints refer to models where coupling to charm mass is forbidden at tree level, while dotted lines show the constraints for models where it is allowed.  
		Note how the MICROSCOPE bound becomes mass-dependent, unlike the linear case.
		Above the dashed yellow line, sourcing effects cannot be neglected, see main text. 
}
	\label{fig:QuadBounds}
\end{figure}

Here, we collect the bounds on the two models studied in this work, and compare them to the estimated bounds at flavour factories. They are shown in \cref{fig:LinBounds,fig:QuadBounds} for the linear and quadratic model, respectively.

For the linear model, we perform a parameter scan to find $V_d, \theta_{0}$ such that the experimental values of the modulus of each CKM matrix element, as well as the Jarkslog invariant, is reproduced within two standard deviation.
We then recast the MICROSCOPE mission~\cite{MICROSCOPE:2022doy} on EP, using the prescription of~\cite{Damour:2010rp}, taking into account the strange mass dependence of the nucleon masses as well~\cite{Junnarkar:2013ac}.
We also use the bounds on the light quark mass variation put using atomic clocks and \ce{^{229}Th} spectroscopy~\cite{Arakawa:2026mls}.
For flavour factories, we show the reach of NA62, as well as the b-factories LHCb and BelleII, using the $2\sqrt{N_{D}}|V_{cs}|/|V_{cd}|$ sensitivity advocated for in~\cite{Dine:2024bxv}.

As mentioned in \cref{sec:model}, the quadratic model induces a $\theta^2 G^2$ coupling at tree level.
Just like the linear model, this is constrained by using atomic clocks and nuclear spectroscopy~\cite{Arakawa:2026mls}. 
Moreover, it induces violation of the equivalence principle, constrained from the MICROSCOPE mission~\cite{MICROSCOPE:2022doy}.
Indeed, while in the linear case the EP bound comes from tree-level exchange of the light boson, in the quadratic case, the DM profile around Earth induces EP violations which are proportional to $\theta^2$~\cite{Hees:2018fpg}.
Unlike the linear coupling, the quadratic one induces non-oscillating changes in fundamental constants, which are strongly constrained by Big-Bang Nucleosynthesis~\cite{Coc:2006sx,Blum:2014vsa,Stadnik:2015kia,Sibiryakov:2020eir,Bouley:2022eer}, however these constraints strongly depend on the cosmological history of $\theta$, and are thus not shown.
To avoid clutter, we likewise do not show limits coming from pulsar timing arrays~\cite{Gan:2025icr}, that are weaker than the ones set by atomic clocks.
Depending on the sign of the coupling, sourcing of $\theta$ due to matter density, both on Earth and in astrophysical objects~\cite{Hook:2017psm,Hees:2018fpg,Balkin:2020dsr,Zhang:2021mks,Balkin:2022qer,Banerjee:2022sqg,Bauer:2024hfv,Balkin:2023xtr,Bauer:2024yow,Banerjee:2025dlo,delCastillo:2025rbr}, can invalidate the constraints coming from experiments on Earth. 
We estimate the critical coupling by~\cite{Hees:2018fpg}:
\begin{equation}
	\frac{\epsilon^2\lambda}{y_c f_{\rm crit}^2} = \frac{R_{E}}{3\alpha_{E}^g M_{E}}\,,
	\label{eq:CritCoupling}
\end{equation}
with $\alpha_E^g \approx 1$ the gluon dilatonic charge of Earth.
Above this coupling, the estimates for the limits are much more delicate, see~\cite{Banerjee:2025dlo,delCastillo:2025rbr}.

At this point, it should be clear that the stringent bounds on this model are due to the coupling to the charm at tree level.
Importantly, the estimated sensitivity of flavour factories is not affected by this coupling.
If one were to forbid it at tree level, all existing bounds would be suppressed not only by loop suppression, but also by the quark mass running in the loop, just like in the linear model.
This greatly reduces the effective coupling to gluons, leaving however a sizable coupling to light quark masses.
The constraints and estimated sensitivity are show in~\cref{fig:QuadBounds} in solid lines, while the constraints coming from tree-level coupling to charm are shown in dotted lines.


We also comment on the possibility of having $\theta^2$ couple to down-type quarks $\mathcal{L} \supset\lambda \epsilon^2 \theta^2 \bar{Q}_1 H s_R$, which contains a contribution to the strange mass, instead of the charm mass, suppressed by $\lambda$.
The NA62 bound is also enhanced in this case, by $\sqrt{m_c/m_s}$.
Importantly, the strange quark contribution is the smallest out of all contributions, being smaller than the other quarks by $\sim 30\%$~\cite{Hoferichter:2025ubp}.
This implies that equivalence principle violations, as well as signals in atomic and nuclear clocks are suppressed, and the critical coupling for screening is likewise larger.
Nonetheless, we do not expect these effects to be able to bridge the large gap that can be observed in~\cref{fig:QuadBounds}, left.
If tree-level $\phi^2 \bar s s$ is forbidden, then the constraints are expected to look similar as~\cref{fig:QuadBounds}, right, since the non-flavour probes are sensitive to coupling to light quark masses. 
The latter are generated in the same way as if the coupling is to up-type quarks, except with the same enhancement that the flavour probe enjoy.

\section{Conclusions}
\label{sec:Conclusions}
We have studied the possibility that ultralight dark matter induces time-dependent oscillations in the CKM matrix. Such effects arise in frameworks in which a light scalar field couples to the flavor sector.
In particular, we have revisited the linearly coupled scenario based on the Nelson--Barr mechanism~\cite{Dine:2024bxv}, and further consider models featuring quadratic couplings to Standard Model operators, by motivating the existence of a technically-natural effective Froggatt--Nielsen model where the quadratic coupling dominates, and the linear couplings are suppressed.
While EP violation bounds are not completely avoided in the quadratic model, one obtains an ameliorated bound at larger masses with respect to the linear case.

The main phenomenological consequence is a periodic modulation of flavor observables, which can be interpreted as a time-dependent effective lifetime. 
In turn, the latter can be uncovered using time-resolved measurements, whereas time-averaged measurements have a severely reduced, albeit nonzero, sensitivity.
We performed a detailed statistical analysis of different observables, focusing on the parameters of the NA62 experiment, and showed that lifetime-based measurements suffer from a nontrivial degeneracy when the total number of produced particles is not precisely known.
Namely, the observation of a finite time window limits the ability to distinguish whether the observed dataset is consistent with a given lifetime and $N_{\rm tot}$ or with a larger (smaller) lifetime and a correspondingly smaller (larger) number of total particles, see \cref{sec:Intuitive}.
In the experimentally relevant regime where only a small fraction of the particle lifetime is observed, this leads to a substantial degradation in sensitivity.
If $N_{\rm tot}$ is known up to some a finite uncertainty, we also show that the scaling with the size of the dataset is nontrivial, and in fact the parametric limitation is always recovered in the asymptotic limit of a large dataset.

Motivated by this, we have explored alternative strategies based on counting observables.
By directly probing the time dependence of the number of events, one can avoid the normalization–lifetime degeneracy and recover the expected statistical scaling.
We find that time-dependent counting measurements, combined with a spectral analysis, offer a viable path to probe ULDM-induced oscillations over a wide range of frequencies.
The main experimental requirement is control over time-dependent systematic at frequencies comparable to the signal. 
In an experiment, typically such systematics will be most prominent for certain frequency ranges and be suppressed for others. 
The expectation is thus that ULDM masses will be probed non-uniformly by one experiment, motivating to pursue this measurement in different set-ups.

More generally, our analysis demonstrates that flavor experiments are sensitive to ULDM through time-dependent effects in the quark sector.
This provides a complementary probe to existing searches based on clocks, equivalence-principle tests, and spin-precession experiments, and motivates further exploration of time-dependent signatures in precision measurements.
\section*{Acknowledgments}

We are grateful for stimulating discussions on aspects of this study with
Fred Blanc,
Augusto Ceccucci,
Akbar Diaz Rodarte,
Lau Gatignon,
Jan Jerhot,
Gaia Lanfranchi,
Francesco Lombardi,
Karim Massri,
Matthew Moulson,
Wolfram Ratzinger,
Giuseppe Ruggiero,
and
Tommaso Spadaro.

The research of PLC is supported by the Max Planck Society--Weizmann Institute of Science joint postdoctoral program.
This work has received funding through the European Research Council under grant ERC-2018-StG-802836 (AxScale) as well as by the Deutsche Forschungsgemeinschaft (DFG, German Research Foundation) under Germany’s Excellence Strategy – EXC 2094 – 390783311 and through DFG Grant No. 532766533.
GP is supported by the Israel Science Foundation (ISF), Minerva, the NSF-BSF, and the European Research Council (ERC, DM-Dawn, Grant Agreement No. 101199868).
\bibliographystyle{JHEP}
\bibliography{Inspire, nonInspire}
\appendix
\section{Parameter Estimation}
\label{sec:MLE}
The estimation of a parameter from experimental data is a classical problem in statistics.
In this paper, we use maximum-likelihood estimation (MLE).
Given a data set, where each observation $x_i$ is assumed to be drawn from a probability distribution $p(x|\Gamma)$, one defines the likelihood function
\begin{equation}
	L = \prod_i p(x_i|\Gamma)
	\label{}
\end{equation}
It is then useful to work with the logarithm of the likelihood, so that the product turns into a sum.
Also, since a parameter-independent normalization would not affect the procedure, we always define the log-likelihood up to constant terms in $\Gamma$.
Assuming the dataset is made up of independently drawn observations and is asymptotically large, maximizing the log-likelihood with respect to $\Gamma$ gives an unbiased, consistent and equivariant estimate that is optimal in the sense that it has minimum uncertainty.

Explicitly, the MLE estimate $\Gamma^{\rm MLE}$ is defined as the value where
\begin{equation}
\left. {\frac{\rm{d} \ln L}{\rm{d} \Gamma}}\right|_{\Gamma = \Gamma^{{\rm MLE}}}\mkern-50mu=0\, .  \label{<+label+>}
\end{equation}
The expected variance $\sigma^2_\Gamma$ of the estimator $\Gamma^{\rm MLE}$ can be computed in terms of the curvature of the log-likelihood at $\Gamma = \Gamma^{\rm MLE}$, known as the Fisher information:
\begin{align}
	\mathcal{I} &=-\mathds{E}[\left. \frac{\partial^2\ln L(\{x_i\}_i|\Gamma)}{\partial \Gamma^2}\right|_{\Gamma=\Gamma_{\rm MLE}}\mkern-50mu] \,,\\
		\sigma &= \frac{1}{\sqrt{\mathcal{I}}} \, ,
	\label{eq:MLEUnc}
\end{align}
where $\mathds{E}$ denotes the expected value.
The Cramer-Rao bound states that the variance of any unbiased estimator is bounded from below by the MLE variance in \cref{eq:MLEUnc}, and the MLE estimate is unbiased in the asymptotic limit of infinite and independent data points.
Hence, parameter estimation using the MLE optimal in the statistical sense.
\section{An Intuitive Derivation of the Uncertainty in the Unknown $N_{\rm tot}$ Case}
\label{sec:Intuitive}
In the limit where $\Delta y \ll 1$, the exponential distribution can be reduced to a linear distribution $1 - \Gamma t$ in the finite interval $t \in [T_1, T_2]$.
Then, the relative standard error on the best estimator for $\Gamma$ is:
\begin{equation}
\frac{\sigma_\Gamma}{\Gamma} = \frac{1}{\Gamma \sqrt{N} \sigma_t} \, .
	\label{<+label+>}
\end{equation}

The variance of $t$ can easily be computed, and by expanding in $\Delta y \ll 1 $ we obtain:
\begin{equation}
	\sigma^2_t = T^2(\frac{1}{12} -\frac{\Delta y^2}{144} + \mathcal{O}(\Delta y^3)) \, ,
	\label{}
\end{equation}
here, the first contribution is simply the variance of a flat distribution.
If we drop all $\Delta y$-suppressed correction, we find the upper line of \cref{eq:approxSigma}.
This simple argument reproduces the result of the rigorous MLE, as well as explaining the factor $2\sqrt{3}$.

\begin{figure}[htpb]
	\centering
	\includegraphics[width=0.8\textwidth]{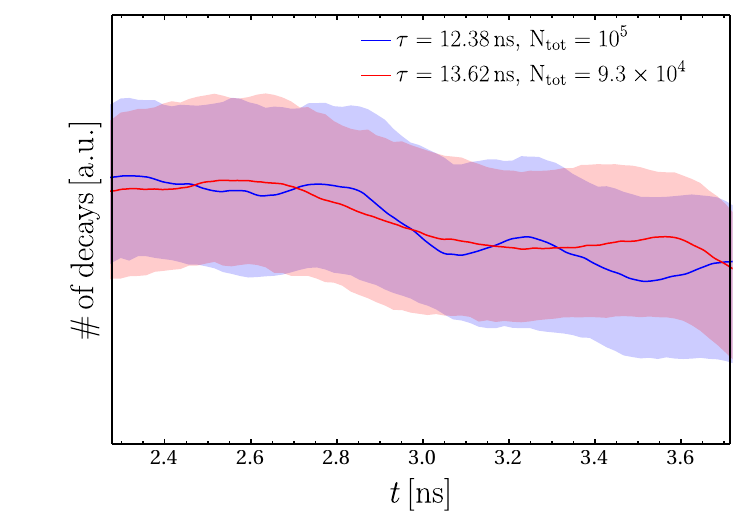}
	\caption{Best non-parametric fit and 95\% confidence bands of two toy dataset drawn with different normalization and lifetime, chosen such that the number of decays in the observed region is the same. The relative difference in lifetime for the two datasets is one order of magnitude larger than $1/\sqrt{N_{\rm obs}}\approx 10^{-2}$, but the two fits are statistically indistinguishable, due to the degeneracy between normalization and lifetime.}
	\label{fig:Deg}
\end{figure}

We can think of this factor as accounting for the degeneracy that exists between the total normalization and the lifetime in this regime: if $\Delta y$ is small, it is hard to distinguish between distributions with different $\Gamma,N_{\rm tot}$, provided they have the same $N_{\rm obs}$, see \cref{fig:Deg}.

\section{Lifetime Uncertainty when Treating $N_{\rm tot}$ as a Nuisance Parameter}
\label{app:LifeUnc}
In this appendix, we confirm the uncertainty estimate in \cref{eq:RelUncLifGauss}.
We use both a frequentist and a Bayesian approach.
In a fully frequentist approach, one treats both $\Gamma$ and $N_{\rm tot}$ as parameters of the likelihood, and a \emph{profile} method is used to get the uncertainty on $\Gamma$ only, while in the hybrid Bayesian approach, the likelihood is \emph{marginalised}, that is the likelihood on $N_{\rm tot}$ is used as a prior and integrated over.

\subsection{Frequentist Approach}
The full likelihood on $\Gamma,\, N_{\rm tot}$ reads:
\begin{equation}
L(\Gamma, N_{\rm tot}) = p(N_{\rm tot}) \cdot L(\{t_i\}, N_{\rm obs}, N_{\rm tot} \mid \Gamma, y_1,y_2) \, ,
\label{eq:likelihood}
\end{equation}
which needs to be maximised with respect to both its parameters.
To profile over $N_{\rm tot}$, one considers, for each value of $\Gamma$, the value of $N_{\rm tot}$ that maximizes the likelihood.
This produces a \emph{profiled likelihood} $L_\mathrm{profile}(\Gamma)$, which depends only on $\Gamma$.
The $1\sigma$ uncertainty on $\Gamma$ is then extracted from the profile likelihood in the following way.
One first computes the two-dimensional log-likelihood $\ln L(\Gamma, N_{\rm tot})$ and identifies its global maximum $(\Gamma^{\rm MLE}, N_{\rm tot}^{\rm MLE})$.
Around this maximum, the $1\sigma$ contour is defined by
\begin{equation}
	\ln L(\Gamma, N_{\rm tot}) - \ln L(\Gamma^{\rm MLE}, N_{\rm tot}^{\rm MLE}) = -\frac{1}{2}\,.
\label{eq:contour}
\end{equation}
The intersection of this contour with the line where 
\begin{equation}
\frac{\partial \ln L}{\partial N_{\rm tot}} = 0 \, ,
	\label{}
\end{equation}
(i.e., the value of $N_{\rm tot}$ that maximizes the likelihood for each $\Gamma$) determines the uncertainty of $\Gamma$ as the difference between the two corresponding $\Gamma$ values.
In simpler terms, profiling ``slides'' the nuisance parameter $N_{\rm tot}$ to its best-fit value for each $\Gamma$, and the resulting one-dimensional likelihood curve in $\Gamma$ provides its uncertainty.
We have confirmed by explicit toy simulations that the uncertainty obtained via this method is consistent with \cref{eq:RelUncLifGauss}.
\subsection{Bayesian Approach}
In a Bayesian approach, one would marginalize over $N_{\rm tot}$ in the likelihood.
The marginalised likelihood is:
\begin{equation}
	L(\Gamma) = \sum_{N_{\rm tot}\geq N_{\rm obs}} p(N_{\rm tot})  
	\begin{pmatrix}N_{\rm tot}\\N_{\rm obs}\end{pmatrix}
	L(\{t_i\}_i,N_{\rm obs},N_{\rm tot}|\Gamma, y_1,y_2)  \, ,
	\label{eq:MarginalisedL}
\end{equation}
where we had to introduce a binomial factor for combinatorial reasons, which can contribute if $p(N_{\rm tot})$ is nontrivial.
Since we use a gaussian prior on $N_{\rm tot}$, we employ a saddle point approximation, that is we seek an approximation to the quantity
\begin{equation}
F=\ln\sum_m \frac{1}{\sqrt{2\pi \sigma^2}}\exp\left( \frac{(\mu- m)^2}{2\sigma^2} \right)
	\begin{pmatrix}N_{\rm obs} + m\\N_{\rm obs}\end{pmatrix}
	\exp(-m\Gamma T)\,,
	\label{<+label+>}
\end{equation}
where $m\equiv N_{\rm tot}- N_{\rm obs}$.
We begin by approximating the sum with an integral, since we expect $N_{\rm obs}, \, N_{\rm tot} \gg 1$, and the exponential to vary little between integers around $\mu$. 
\begin{equation}
	F \approx \ln\int_m f(m,\Gamma)\,.
	\label{<+label+>}
\end{equation}
Next, we can use the Laplace approximation for the integral:
\begin{equation}
	F\approx f(\bar{m},\Gamma) -\frac{1}{2}\left(\ln\left( \frac{\partial^2f(m,\Gamma)}{\partial m^2}\right){\Bigg|}_{m = \bar{m}}
	\mkern-35mu - \ln(2\pi)\right) \,,
	\label{<+label+>}
\end{equation}
where $\bar{m}=\bar{m}(\Gamma)$ solves $\frac{\partial f}{\partial m}(\bar{m},\Gamma) = 0$ and its existence as a smooth function of $\Gamma$ is guaranteed by the implicit function theorem.
We can now compute the Fisher information as the second derivative with respect to $\Gamma$ of $F$.
In the process, we need the first two derivatives of $\bar{m}(\Gamma)$. 
Using again the implicit function theorem, and imposing that $\bar{m}$ is a critical point of $F$, one obtains
\begin{equation}
	\bar{m}_{\Gamma} = - \frac{f_{m\Gamma}}{f_{mm}} \, ,
	\label{}
\end{equation}
where we switched to Lagrange notation for the derivatives. 
Simple use of the chain rule then allows to obtain the Fisher information, whose explicit form is not particularly illuminating so we refrain from reporting it here.
Following this approach, we again find an expression which is both qualitatively and quantitatively in agreement with \cref{eq:RelUncLifGauss}.

%

\end{document}